# A Non-Volatile Cryogenic Random-Access Memory Based on the Quantum Anomalous Hall Effect

Shamiul Alam[1*], Md Shafayat Hossain[2*], and Ahmedullah Aziz[1]

[1]Dept. of Electrical Eng. & Computer Sci., University of Tennessee, Knoxville, TN, 37996, USA
[2]Dept. of Electrical Engineering, Princeton University, Princeton, NJ, 08544, USA
*These authors have contributed equally

**Abstract**

**The interplay between ferromagnetism and topological properties of electronic band structures leads to a precise quantization of Hall resistance without any external magnetic field. This so-called quantum anomalous Hall effect (QAHE) is born out of topological correlations, and is oblivious of low-sample quality. It was envisioned to lead towards dissipationless and topologically protected electronics. However, no clear framework of how to design such an electronic device out of it exists. Here we construct an ultra-low power, non-volatile, cryogenic memory architecture leveraging the QAHE phenomenon. Our design promises orders of magnitude lower cell area compared with the state-of-the-art cryogenic memory technologies. We harness the fundamentally quantized Hall resistance levels in moiré graphene heterostructures to store non-volatile binary bits (1, 0). We perform the memory write operation through controlled hysteretic switching between the quantized Hall states, using nano-ampere level currents with opposite polarities. The non-destructive read operation is performed by sensing the polarity of the transverse Hall voltage using a separate pair of terminals. We custom design the memory architecture with a novel sensing mechanism to avoid accidental data corruption, ensure highest memory density and minimize array leakage power. Our design is transferrable to any material platform exhibiting QAHE, and provides a pathway towards realizing topologically protected memory devices.**

Electronic bands of non-trivial topology give rise to quantum phenomena at macroscopic scale. Under a strong magnetic field, two-dimensional electron systems exhibit such a quantum state- the quantum Hall effect, i.e. the quantization of Hall resistance. Back in 1988, Haldane[2] predicted that a quantum Hall state can arise even in the absence of an external magnetic field. This new state became known as the quantum anomalous Hall effect (QAHE). However, it took more than two decades for the first experimental realization of QAHE, when it was demonstrated in a magnetic topological insulator [Cr-doped Bi(Sb)$_2$Te$_3$][3–10]. The experimental fingerprint of QAHE is straightforward: The Hall resistance is quantized to h/$v$e$^2$ (where $v$, the so-called Chern number, is an integer that depends on the topological properties of the band structure) at zero magnetic field. Such simple and universal values stem from the current being carried by lossless edge channels which require neither high electron mobility [11,12] nor external magnetic fields. Moreover, on a given edge, the current flows only in one direction. Thanks to such transport properties which are also immune to sample complexities [11,12], QAHE can potentially be useful in spin-filtering[11], resistance metrology [13] and topological quantum computing [14]. However, how we can materialize these prospects and come up with a QAHE-based electronic device have remained elusive so far.

Here we bridge the gap between the QAHE physics and device architecture to build the framework of a scalable non-volatile memory. We utilize the quantization of Hall resistance in QAHE to design a memory cell and then construct a 3D cross-point memory array capable of efficient read and write operations.

To design our memory cell, we use the QAHE reported in a twisted bilayer graphene (tBLG) on hexagonal boron nitride (hBN) moiré heterostructure, where the quantization of Hall resistance persists at temperatures as high as 6 K [1]. This is significantly higher compared to the earlier reports on transition metal doped (Bi, Sb)$_2$Te$_3$ thin films [15–18]. With the ongoing thrust in material discoveries, room-temperature applications will be feasible. Very recent works on QAHE in TbMn$_6$Sn$_6$, MnBi$_2$Te$_4$ offer such prospects [19–21]. Importantly, our design can easily be implemented with any quantum materials, allowing a wide-range of applications.

We start by delineating the dynamics of a QAH insulator and how it can be used as a non-volatile memory cell. Fig. 1(a) illustrates the schematic of a QAH insulator with electric contacts to apply bias current and to measure both longitudinal and transverse (Hall) voltages. The QAH states occur when an applied gate voltage tunes the bulk carrier density to close to zero. Then, the Hall resistance shows a hysteretic switching with the change of magnetic field. Also, magnetic domains of the ferromagnetic materials (for example, tBLG) strongly interact with the applied electrical current, which provides the electrical control over the polarization of the magnetic domains [22]. It has also been shown [1] that the switching of Hall resistance driven by DC bias current is similar to that driven by magnetic field. Fig. 1(b) shows a schematic of the hysteretic switching of Hall resistance $R_{xy}$ (=$V_{xy}/I_{Bias}$) with the bias current at absolutely zero external magnetic field. Here, we mark two critical values of bias current, $I_{C-}$ and $I_{C+}$, which denote the values of bias current required for the switching of Hall resistance between $\pm h/e^2$.

Figure 2 captures the highlight of our idea to leverage the quantized Hall resistance of a QAH insulator as a non-volatile memory cell along with the write and read operations. We define the quantized Hall resistances, $-h/e^2$ and $+h/e^2$, as logic '0' and logic '1' respectively. Based on this definition, $I_{Bias}$ can be divided into three regions [Fig. 2(a)]- (i) $I_{Bias} \leq I_{C-}$: write '0' region, (ii) $I_{Bias} \geq I_{C+}$: write '1' region, and (iii) $I_{C-} < I_{Bias} < I_{C+}$: read region. Fig. 2(b) shows the bias current dependence of the Hall voltage, $V_{xy}$ (=$I_{Bias} \times R_{xy}$). Importantly, as seen in Fig. 2(b), write operations of both logic '0' and '1' entail positive $V_{xy}$ because $I_{Bias}$ and R$_{xy}$ are either both positive [when writing logic '1'; Fig. 2 (c)], or both negative [when writing logic '0'; Fig. 2 (d)]. In sharp contrast, read operations of logic '0' and '1' manifest opposite sign of $V_{xy}$. For example, if we use a positive bias current ($0 < I_{Bias} < I_{C+}$), logic '1' and logic '0' correspond to positive and negative $V_{xy}$, respectively. The clear difference in the sign of the Hall voltage for logic '0' and '1' makes the sensing of the memory states simple and straightforward. Fig. 2(g) summarizes the key idea of a non-volatile memory utilizing the quantization of the Hall resistance in a QAH insulator, listing $I_{Bias}$ for write and read operations, as well as the state of the Hall resistance and Hall voltage for the two memory states.

It is worthwhile to analyze the operation of the proposed QAHE based non-volatile memory in a 3D cross-point memory array. Fig. 3(a) schematically shows our memory element: a tBLG moiré heterostructure where tBLG (with interlayer twisted angle of 1.1°) is encapsulated between flakes of hBN [14]. We have developed a Verilog A based phenomenological model for the QAHE in tBLG moiré heterostructure to use in our analysis of the heterostructure as a memory cell. The model can also be calibrated for other QAH insulators using the values of bias currents ($I_{C-}$ and $I_{C+}$) and the Chern number ($\nu$). A behavioral representation of our model is shown in Fig. 3(b). Fig. 3(c) shows $R_{xy}$ as a function of $I_{Bias}$ at zero external magnetic field and at T = 4 K which is obtained from the model shown in Fig. 3(b). As seen from Fig. 3(c), we can deduce the values of $I_{C-}$ and $I_{C+}$ for the tBLG heterostructure (approximately -4 nA and 100 pA, respectively) which are crucial for the memory operation. In our phenomenological model, we also account for the temperature dependence of Hall resistance using the following equation [1]:

$$R_{xy} = \frac{h}{e^2} - R_1 e^{-\frac{\Delta}{T}}, \tag{1}$$

where, $R_1$ is a fitting constant, $\Delta$ (= 26 ± 4 K) is the energy required to create and separate an excitation of particle-antiparticle of the QAH state and T is the temperature. Fig. 3(d), which shows the temperature dependence of the Hall resistance obtained from our model, agrees reasonably with the measured values reported in Ref. [1].

To evaluate the practicality of a memory device, it is crucial to consider an array level scenario. The unique properties of our QAHE based memory device necessitate a custom designed memory array. To ensure the highest storage density, we adopt the cross-point memory architecture [23] with slight modification in the inter-cell connection pattern. A major component of the cross-point array is a two-terminal selector device [24] that allows access to a specific cell for read/write operation and suppresses current flow through the other cells. The selector devices are connected in series with the memory elements in every cell. A multitude of selector devices exist [25–30] with diverse selectivity (ON/OFF ratio) and switching thresholds. For our QAHE based memory, the Cu-containing mixed-ionic-electronic-conduction (MIEC) material will be a perfect selector, due to its high selectivity (~$10^6$), and ultra-low leakage (< 10 pA) [31–33]. We have developed a look-up-table based phenomenological model (in Verilog-A) for this selector and calibrated the model [Fig. 3(e)] with the measured current-voltage (*I-V*) characteristics reported in Refs. [31–33].

Fig. 3(f) shows the schematic view of our proposed array architecture. Every memory cell (QAHE device + selector) is sandwiched between orthogonally running metal lines, namely word line (WL) and bit line (BL). The WLs and the BLs are shared along the rows and columns of the array, respectively. The difference between the DC voltages applied on the WL and BL of a particular a memory cell essentially dictates the effective voltage across the cell ($V_{Cell}$). Note, the substrate terminal of every cell is biased with a constant voltage to maintain an appropriate electron density [not illustrated in the figures for simplicity]. Conveniently, we need only one control input, $V_{Cell}$, to read from or write into the QAHE devices. We utilize three different levels of $V_{Cell}$ (with appropriate polarity) to write logic '1/0' and to read the stored data.

Our architecture entails accessing one cell per row at a time. The Hall-voltage terminals of the neighboring cells in the same row are electrically connected in series with each other [Fig. 3(f)]. Importantly, the algebraic sum of the Hall voltages of all the cells in a row ($V_{hr-i} = \sum_{j=1}^{n} V_{xy-ij}$) bears the signature of the stored data in the accessed memory cell, along with some residual contributions from the other cells in the same row. For example, consider that the $k^{th}$ cell of the $i^{th}$ row is being accessed. The sum of the Hall voltage for this row can be expressed as:

$$V_{hr-i} = \sum_{j=1}^{n} V_{xy-ij} = I_{\text{Bias}-ik} \times R_{xy-ik} + \sum_{j=1,\ j \neq k}^{n} I_{\text{Bias}-ij} \times R_{xy-ij}. \tag{2}$$

The bias current ($I_{\text{Bias}}$) through the accessed cell is usually orders of magnitude higher [see Fig. 4(k)] than the leakage current through the other cells [half-accessed row (HAR) cells] in the same row; $I_{\text{Bias-HAR}} \ll I_{\text{Bias-accessed}}$. As a result, the polarity of $V_{hr}$ is dictated by the $R_{xy}$ of the accessed cell. For the tBLG moiré heterostructure based QAHE device, we need to amplify the $V_{hr}$ from a few tens of microvolts to tens of millivolts using cryogenic amplifiers. A suitable candidate for such an amplifier can be one of the cryogenic low-noise amplifiers reported in Refs. [34–36]. After the amplification, the millivolt level Hall voltage is used to determine the memory state of the accessed cell. Recall, during read operation, logic '0' and '1' memory states correspond to opposite polarity of Hall voltage [Fig. 2(b)]. Therefore, we feed the amplified $V_{hr}$ to a cryogenic voltage comparator [37] to sense the memory state of the accessed cell.

In Fig. 4, we present the simulated memory operations (read/write) in our proposed QAHE based cross-point array. Fig. 4(a) illustrates four types of memory cells in a block of cross-point array. We utilize a standard biasing scheme for cross-pint arrays, commonly known as the *V/2* biasing [30]. Different levels of access voltage ($V_{ACC}$) are applied to read from or write into a specific cell. The row (column) that holds the accessed cell is called the half-accessed row (column), because the inactive cells in this row (column) receive half of the access voltage ($V_{ACC}/2$). All the other unaccessed cells ideally have zero voltage across them. Without any loss of generality, we assume $M_{11}$ [Fig. 3 (f)] to be the accessed cell and examine its read/write dynamics. Fig. 4(b) shows the cell current levels through the accessed, half-accessed, and unaccessed cells during 1→0 memory write operation. Only the accessed cell exhibits a transition (+$h/e^2$ → -$h/e^2$) in the Hall resistance ($R_{xy}$) [Fig. 4(c)], indicating a successful write operation in the accessed cell without disturbing the other cells. Figures 4(d) and 4(e) show similar time dynamics for the 0→1 write operation. Note, the cell current level for the 0→1 operation is of the opposite polarity compared to the 1→0 operation [Fig. 4(d)]. The repeated write operations (0→0 and 1→1) have also been tested and are presented in supplementary Fig. S2.

Next, we examine the read dynamics [Figs. 4(f)-4(j)] of the proposed memory cell. The current flow direction for the read operation is the same as the 1→0 write operation [Fig. 4(f)]. However, the magnitude of the read current is about one-third of the critical current of transition ($I_{C-}$), which alleviates the possibility of accidental 1→0 data flip. In addition, the chosen polarity of the read current nullifies the possibility of accidental 0→1 transition. The read current through the QAHE device generates a Hall voltage ($V_{xy}$), whose polarity is dictated by the previously set Hall resistance ((+$h/e^2$ or -$h/e^2$). In our design, positive/negative polarity of $V_{xy}$ corresponds to logic 0/1, respectively [Fig. 4(g), 4(i)]. It is worth noting, the read operation utilizes different pair of terminals than the write operation, allowing independent optimization of read/write peripheral circuits. Furthermore, the amplifier-comparator pairs are enabled only during the read operation to minimize peripheral power demand. The comparator in our design also provides digital outputs corresponding to the stored data in the accessed cell [Figs. 4(h) & 4(j)]. Also, the bias currents that flow through the cells during different memory operations are shown in Fig. 4(k) which clearly demonstrate that the memory states of the half-accessed and unaccessed cells will not be disturbed during write or read operation in the accessed cell. Thus, our architecture lays out a device-to-array design pathway for the QAHE based unique memory devices.

We close by discussing the broad impact of our work. This manuscript makes the first mark to construct a working electronic device using topological properties of materials. Here we leverage QAHE, that does not require an external magnetic field, and design an elegant memory device that can be built with topological quantum materials. The proposed non-volatile memory architecture is a strong candidate for the cryogenic memory system, thanks to its ultra-low temperature compatibility. Note, the cryogenic memory block is a crucial component of quantum computing systems based on superconducting qubits [38]. Our proposed QAHE-based memory device offers at least a million times reduction in the cell area and 1000 times reduction in the cell read/write power compared with the state-of-the-art cryogenic memory devices [39–46] (see supplementary Table S1 for detailed comparison). Our proposed memory device is a potential game-changer for scalable quantum computing systems [38] and space cryogenics [47].

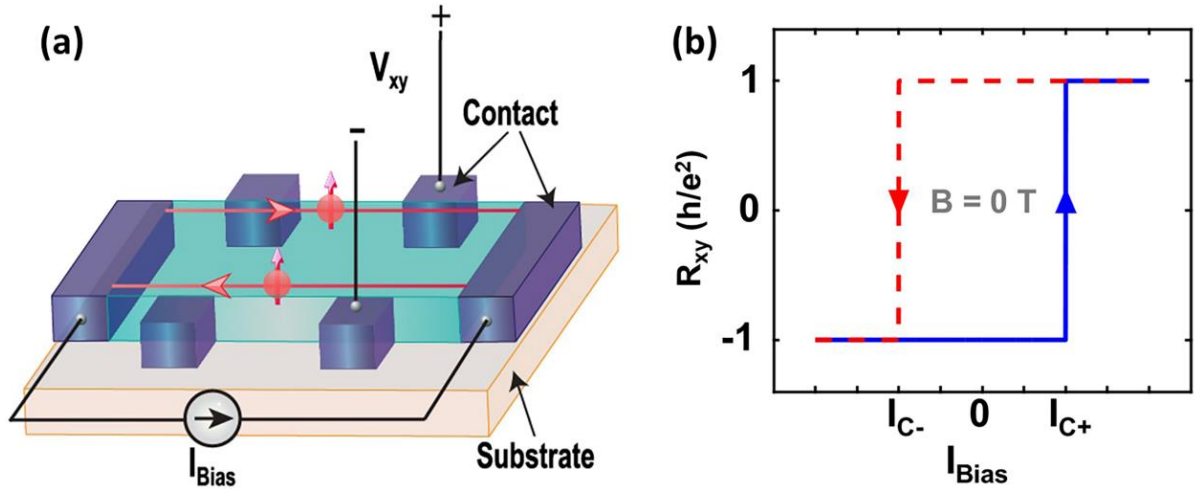

**Fig. 1: Schematic representation of quantum anomalous Hall effect. (a)** Schematic of our device, where the red circles represent the electron and the red arrows show the spin of the electrons. A gate voltage can be applied at the back of the substrate to control the electronic density. $V_{xx}$ and $V_{xy}$ are the longitudinal and transverse (Hall) voltages, respectively, developed in response to the bias current, $I_{Bias}$. **(b)** Illustration of the Hall resistance $R_{xy}$ vs. the bias current, $I_{Bias}$ at zero external magnetic field. $I_{C-}$ and $I_{C+}$ are two critical values of $I_{Bias}$ which determine the hysteretic switching of $R_{xy}$ between $-h/e^2$ and $h/e^2$.

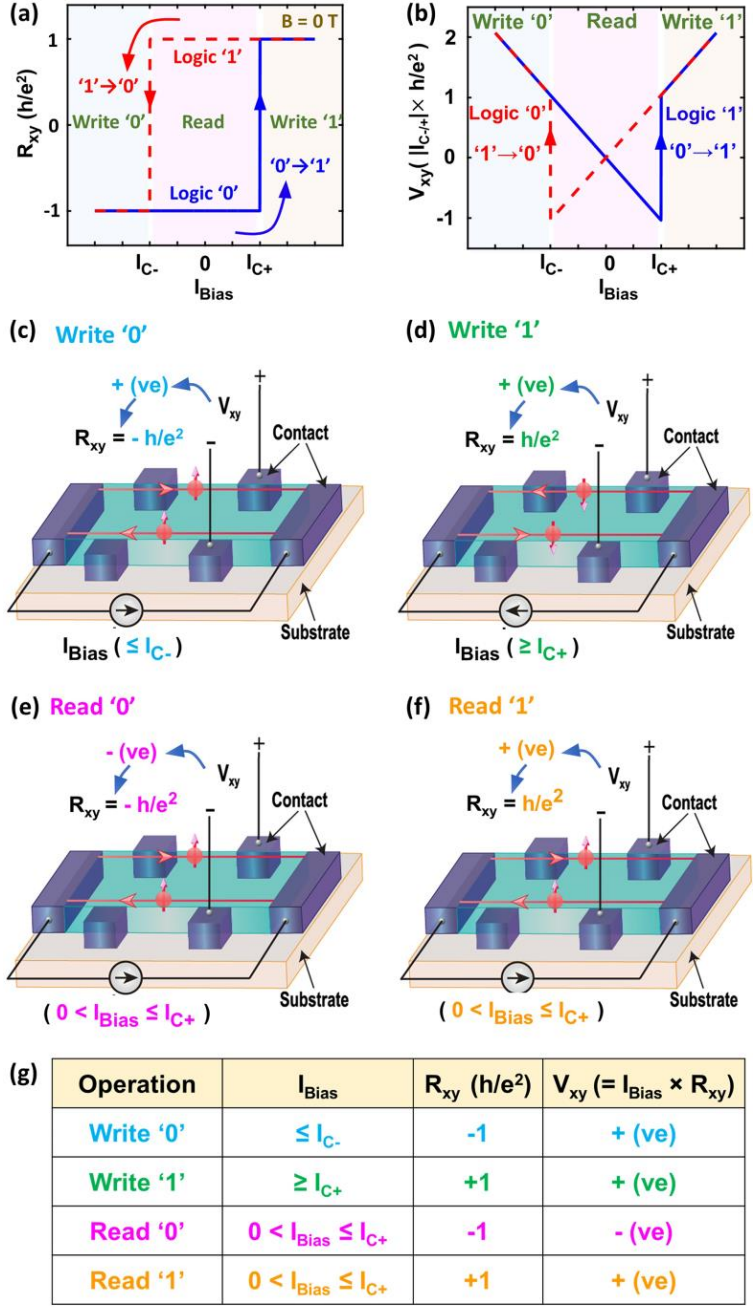

**Fig. 2: Illustration of the quantum anomalous Hall effect based memory operation.** (a) $R_{xy}$ vs. $I_{Bias}$ at zero external magnetic field. Hall resistance values of $-h/e^2$ and $+h/e^2$ are defined as logic '0' and logic '1' respectively. Three regions: (i) $I_{Bias} \leq I_{C-}$, (ii) $I_{Bias} \geq I_{C+}$, and (iii) $I_{C-} < I_{Bias} < I_{C+}$ are marked as write '0', write '1' and read region, respectively. We use this division to choose the required $I_{Bias}$ for different memory operations. (b) Hall voltage, $V_{xy}$ plotted as a function of $I_{Bias}$. (c), (d), (e), & (f) The procedures of write '0', write '1', read '0', and read '1' operations, respectively, along with the state of electron spins (red arrows). (g) Summary of the key idea for the QAHE based memory. The table enlists the required ranges of $I_{Bias}$ corresponding to all memory operations, along with the corresponding Hall resistance and Hall voltage levels.

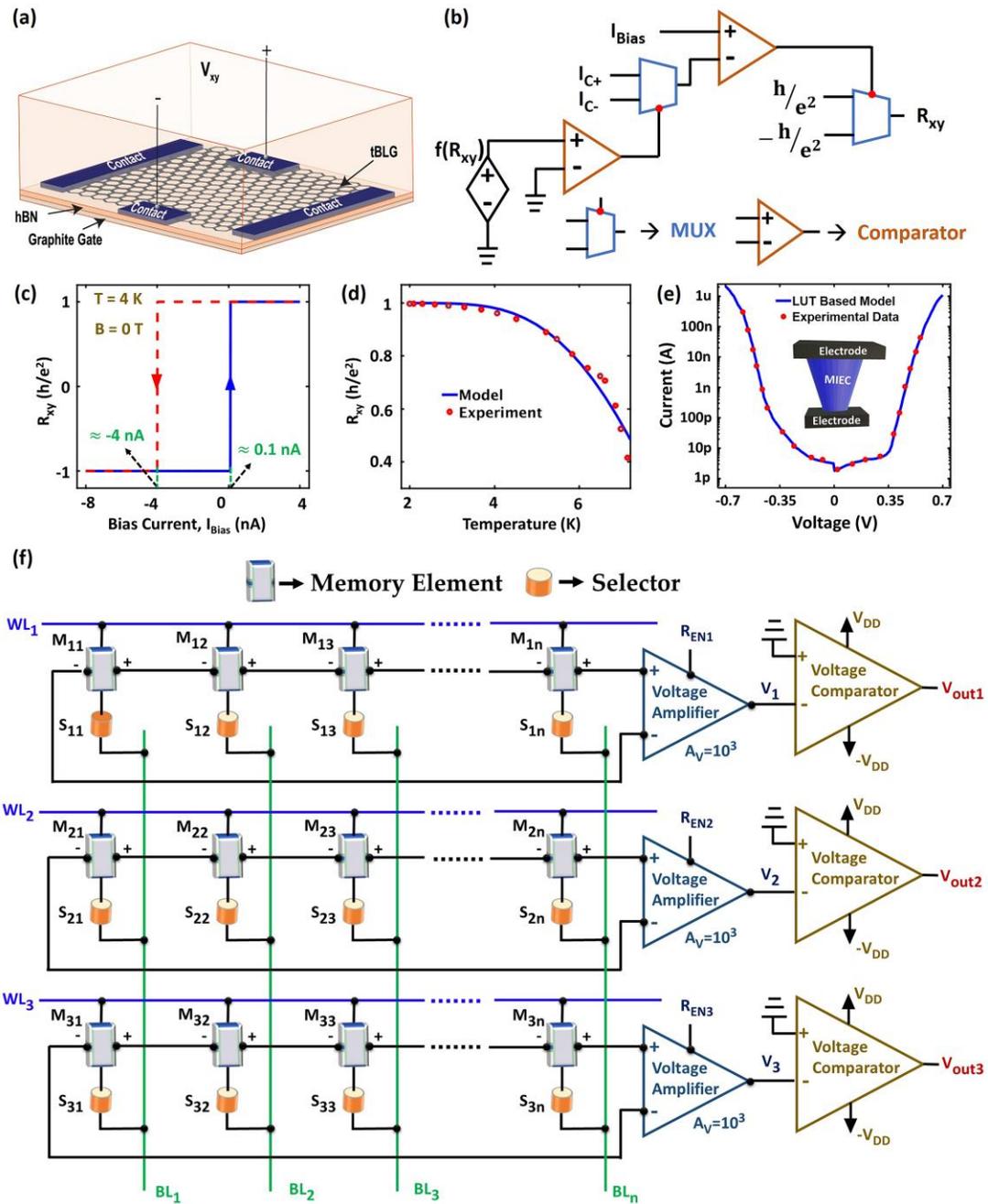

**Fig. 3: Design of the 3D cross-point memory array structure using a tBLG moiré heterostructure as a memory element and mixed-ionic-electronic-conduction based access device as selector.** (**a**) Schematic of a tBLG moiré heterostructure where tBLG is encapsulated between flakes of hBN and a flake of few-layer graphite is used as gate. (**b**) Behavioral representation of our phenomenological model of the observed QAHE in the tBLG moiré heterostructure. (**c**) $R_{xy}$ data, plotted as a function of $I_{Bias}$ at T = 4 K and B = 0 T, exhibit current driven hysteretic switching. (**d**) Temperature dependence of $R_{xy}$ with an experimental matching. (**e**) *I-V* characteristics obtained from a look-up table (LUT) based model of the mixed-ionic-electronic-conduction (MIEC) based selector device. (**f**) Illustration of the overall memory array structure, custom-designed for the QAHE based memory devices.

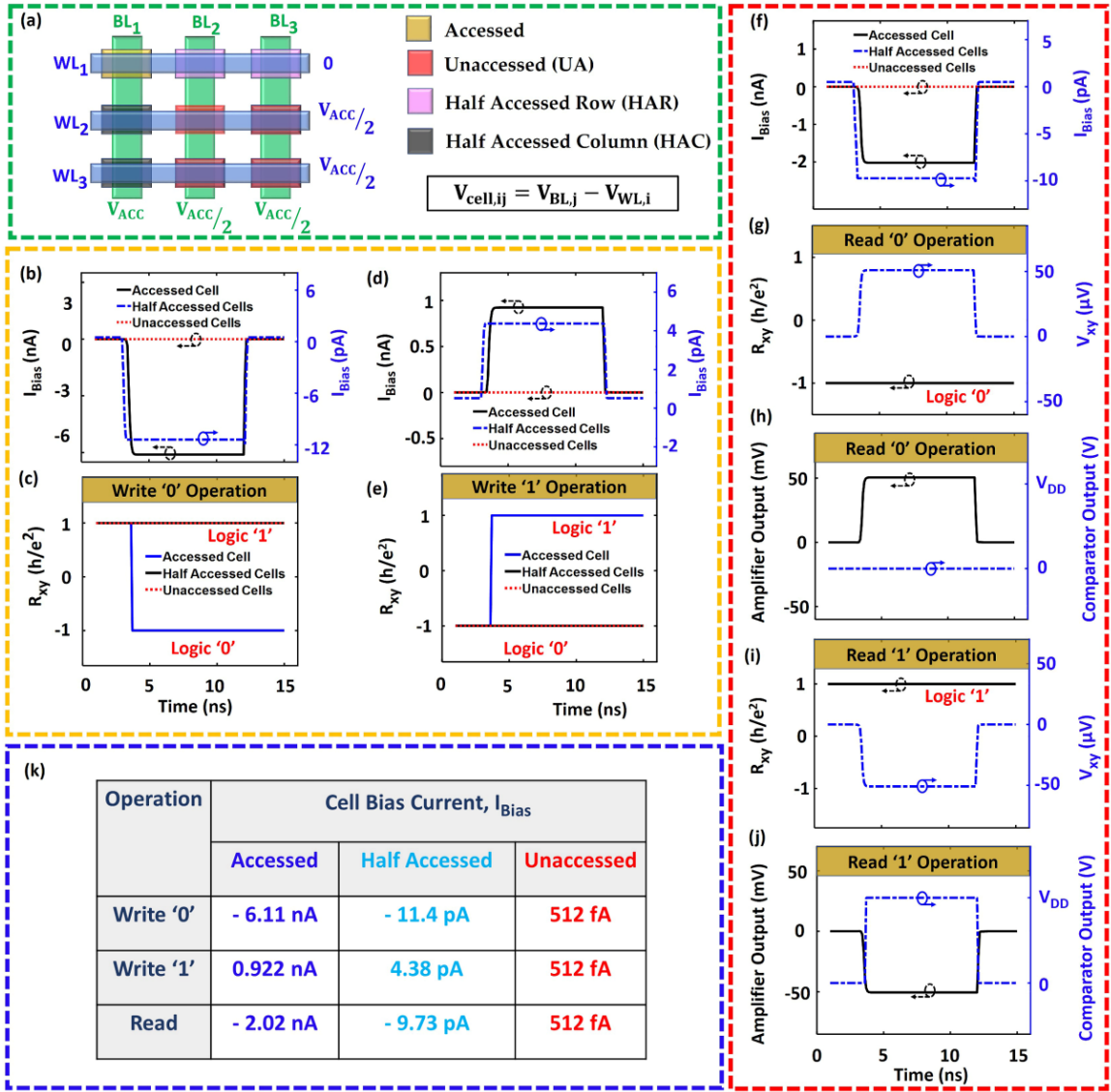

**Fig. 4: Compilation of the simulation results for the memory array.** (a) Arrangement and types of memory cells in a conventional cross-point memory array under the *V/2* biasing scheme. **Write '0' operation:** (b) Cell Bias currents ($I_{Bias}$) through different memory cells generated by applying appropriate $V_{ACC}$ (c) The time dynamics of the Hall resistance ($R_{xy}$) illustrating state transition only in the accessed cell. **Write '1' operation:** (d) $I_{Bias}$ with different magnitude and opposite polarity (compared with the write '0' case) flowing through different memory cells (e) Corresponding transition in $R_{xy}$. **Read operation:** (f) The $I_{Bias}$ is of the same polarity as the write 'o' operation, but has ~3X lower magnitude. The peripheral circuits (amplifier and comparator) are enabled only during this period. The comparator output is always initialized at zero volage. (g) If the memory cell holds binary '0', the Hall voltage ($V_{xy}$) exhibits a positive polarity during the read operation. $R_{xy}$ remains unchanged throughout the process. (h) The amplified $V_{xy}$ is fed into a comparator which produces a digital output. While reading '0', the comparator output remains unchanged at its initial value (0 V). (i) Read '1' operation produces the opposite polarity in $V_{xy}$, and $R_{xy}$ remains unchanged. (j) The amplified $V_{xy}$ triggers a 0→$V_{DD}$ transition in the comparator output, which remains stable throughout the read operation.

**References:**

1. Serlin, M. *et al.* Intrinsic quantized anomalous Hall effect in a moiré heterostructure. *Science (80-. ).* (2020) doi:10.1126/science.aay5533.
2. Haldane, F. D. M. Model for a quantum hall effect without landau levels: Condensed-matter realization of the 'parity anomaly'. *Phys. Rev. Lett.* (1988) doi:10.1103/PhysRevLett.61.2015.
3. Chang, C. Z. *et al.* Experimental observation of the quantum anomalous Hall effect in a magnetic topological Insulator. *Science (80-. ).* (2013) doi:10.1126/science.1234414.
4. Checkelsky, J. G. *et al.* Trajectory of the anomalous Hall effect towards the quantized state in a ferromagnetic topological insulator. *Nat. Phys.* (2014) doi:10.1038/nphys3053.
5. Kou, X. *et al.* Scale-invariant quantum anomalous hall effect in magnetic topological insulators beyond the two-dimensional limit. *Phys. Rev. Lett.* (2014) doi:10.1103/PhysRevLett.113.137201.
6. Bestwick, A. J. *et al.* Precise quantization of the anomalous hall effect near zero magnetic field. *Phys. Rev. Lett.* (2015) doi:10.1103/PhysRevLett.114.187201.
7. Kou, X. *et al.* Metal-to-insulator switching in quantum anomalous Hall states. *Nat. Commun.* (2015) doi:10.1038/ncomms9474.
8. Feng, Y. *et al.* Observation of the Zero Hall Plateau in a Quantum Anomalous Hall Insulator. *Phys. Rev. Lett.* (2015) doi:10.1103/PhysRevLett.115.126801.
9. Chang, C. Z. *et al.* High-precision realization of robust quantum anomalous Hall state in a hard ferromagnetic topological insulator. *Nat. Mater.* (2015) doi:10.1038/nmat4204.
10. Kandala, A., Richardella, A., Kempinger, S., Liu, C. X. & Samarth, N. Giant anisotropic magnetoresistance in a quantum anomalous Hall insulator. *Nat. Commun.* (2015) doi:10.1038/ncomms8434.
11. Oh, S. The complete quantum hall trio. *Science* (2013) doi:10.1126/science.1237215.
12. Jalil, M. B. A., Tan, S. G. & Siu, Z. B. Quantum anomalous Hall effect in topological insulator memory. *J. Appl. Phys.* (2015) doi:10.1063/1.4916999.
13. Götz, M. *et al.* Precision measurement of the quantized anomalous Hall resistance at zero magnetic field. *Appl. Phys. Lett.* (2018) doi:10.1063/1.5009718.
14. Lian, B., Sun, X. Q., Vaezi, A., Qib, X. L. & Zhang, S. C. Topological quantum computation based on chiral Majorana fermions. *Proc. Natl. Acad. Sci. U. S. A.* (2018) doi:10.1073/pnas.1810003115.
15. Lachman, E. O. *et al.* Visualization of superparamagnetic dynamics in magnetic topological insulators. *Sci. Adv.* (2015) doi:10.1126/sciadv.1500740.
16. Lee, I. *et al.* Imaging Dirac-mass disorder from magnetic dopant atoms in the ferromagnetic topological insulator $Cr_x(Bi_{0.1}Sb_{0.9})_{2-x}Te_3$. *Proc. Natl. Acad. Sci. U. S. A.* (2015) doi:10.1073/pnas.1424322112.
17. Wang, W. *et al.* Visualizing ferromagnetic domains in magnetic topological insulators. *APL Mater.* (2015) doi:10.1063/1.4921093.
18. Yasuda, K. *et al.* Quantized chiral edge conduction on domain walls of a magnetic topological insulator. *Science (80-. ).* (2017) doi:10.1126/science.aan5991.
19. Yin, J. X. *et al.* Quantum-limit Chern topological magnetism in $TbMn_6Sn_6$. *Nature* (2020) doi:10.1038/s41586-020-2482-7.
20. Deng, H. *et al.* Observation of high-temperature quantum anomalous Hall regime in intrinsic $MnBi_2Te_4/Bi_2$$Te_3$ superlattice. *Nat. Phys.* 1–7 (2020) doi:10.1038/s41567-020-0998-2.
21. Liu, C. *et al.* Robust axion insulator and Chern insulator phases in a two-dimensional antiferromagnetic topological insulator. *Nat. Mater.* (2020) doi:10.1038/s41563-019-0573-3.
22. Sharpe, A. L. *et al.* Emergent ferromagnetism near three-quarters filling in twisted bilayer graphene. *Science (80-. ).* (2019) doi:10.1126/science.aaw3780.
23. Bez, R. Innovative technologies for high density non-volatile semiconductor memories. in *Microelectronic Engineering* (2005). doi:10.1016/j.mee.2005.04.076.
24. Burr, G. W. *et al.* Access devices for 3D crosspoint memory. *J. Vac. Sci. Technol. B, Nanotechnol. Microelectron. Mater. Process. Meas. Phenom.* (2014) doi:10.1116/1.4889999.
25. Aziz, A., Jao, N., Datta, S. & Gupta, S. K. Analysis of Functional Oxide based Selectors for Cross-Point Memories. *IEEE Trans. Circuits Syst. I Regul. Pap.* (2016) doi:10.1109/TCSI.2016.2620475.
26. Virwani, K. *et al.* Sub-30nm scaling and high-speed operation of fully-confined Access-Devices for 3D crosspoint memory based on mixed-ionic-electronic-conduction (MIEC) materials. in *Technical Digest - International Electron Devices Meeting, IEDM* (2012). doi:10.1109/IEDM.2012.6478967.
27. Hirose, S., Nakayama, A., Niimi, H., Kageyama, K. & Takagi, H. Resistance switching and retention behaviors in polycrystalline La-doped $SrTiO_3$ ceramics chip devices. *J. Appl. Phys.* (2008)

doi:10.1063/1.2975316.
28. Zhang, L. *et al.* High-drive current (>1MA/cm2) and highly nonlinear (>103) TiN/amorphous-Silicon/TiN scalable bidirectional selector with excellent reliability and its variability impact on the 1S1R array performance. in *Technical Digest - International Electron Devices Meeting, IEDM* vols 2015-February 6.8.1-6.8.4 (Institute of Electrical and Electronics Engineers Inc., 2015).
29. Lee, W. *et al.* Varistor-type bidirectional switch (J MAX>10 7A/cm 2, selectivity~10 4) for 3D bipolar resistive memory arrays. in *Digest of Technical Papers - Symposium on VLSI Technology* 37–38 (2012). doi:10.1109/VLSIT.2012.6242449.
30. Aziz, A., Shukla, N., Datta, S. & Gupta, S. K. Implication of hysteretic selector device on the biasing scheme of a cross-point memory array. in *International Conference on Simulation of Semiconductor Processes and Devices, SISPAD* vols 2015-October 425–428 (Institute of Electrical and Electronics Engineers Inc., 2015).
31. Gopalakrishnan, K. *et al.* Highly scalable novel access device based on Mixed Ionic Electronic Conduction (MIEC) materials for high density phase change memory (PCM) arrays. in *Digest of Technical Papers - Symposium on VLSI Technology* (2010). doi:10.1109/VLSIT.2010.5556229.
32. Shenoy, R. S. *et al.* Endurance and scaling trends of novel access-devices for multi-layer crosspoint-memory based on mixed-ionic-electronic-conduction (MIEC) materials. in *Digest of Technical Papers - Symposium on VLSI Technology* (2011).
33. Burr, G. W. *et al.* Large-scale (512kbit) integration of multilayer-ready access-devices based on Mixed-Ionic-Electronic-Conduction (MIEC) at 100% yield. in *Digest of Technical Papers - Symposium on VLSI Technology* (2012). doi:10.1109/VLSIT.2012.6242451.
34. Weinreb, S., Bardin, J. C. & Mani, H. Design of cryogenic SiGe low-noise amplifiers. *IEEE Trans. Microw. Theory Tech.* (2007) doi:10.1109/TMTT.2007.907729.
35. Arakawa, T., Nishihara, Y., Maeda, M., Norimoto, S. & Kobayashi, K. Cryogenic amplifier for shot noise measurement at 20 mK. *Appl. Phys. Lett.* (2013) doi:10.1063/1.4826681.
36. Ivanov, B. I., Trgala, M., Grajcar, M., Ilichev, E. & Meyer, H. G. Cryogenic ultra-low-noise SiGe transistor amplifier. *Rev. Sci. Instrum.* (2011) doi:10.1063/1.3655448.
37. Dziuba, R. F., Field, B. F. & Finnegan, T. F. Cryogenic Voltage Comparator System for 2e/h Measurements. *IEEE Trans. Instrum. Meas.* **23**, 264–267 (1974).
38. Tolpygo, S. K. Superconductor digital electronics: Scalability and energy efficiency issues. *Low Temperature Physics* (2016) doi:10.1063/1.4948618.
39. Yuh, P. F. A 2-kbit Superconducting Memory Chip. *IEEE Trans. Appl. Supercond.* **3**, 3013–3021 (1993).
40. Liu, Q. *et al.* Latency and power measurements on a 64-kb hybrid Josephson-CMOS memory. in *IEEE Transactions on Applied Superconductivity* (2007). doi:10.1109/TASC.2007.898698.
41. Feng, Y. J. *et al.* Josephson-CMOS hybrid memory with ultra-high-speed interface circuit. in *IEEE Transactions on Applied Superconductivity* (2003). doi:10.1109/TASC.2003.813902.
42. Nagasawa, S., Hinode, K., Satoh, T., Kitagawa, Y. & Hidaka, M. Design of all-dc-powered high-speed single flux quantum random access memory based on a pipeline structure for memory cell arrays. *Supercond. Sci. Technol.* (2006) doi:10.1088/0953-2048/19/5/S34.
43. Tahara, S. *et al.* 4-Kbit Josephson Nondestructive ReadOut Ram Operated At 580 psec and 6.7 mW. *IEEE Trans. Magn.* (1991) doi:10.1109/20.133751.
44. Kirichenko, A., Mukhanov, O., Kirichenko, A. F., Mukhanov, O. A. & Brock, D. K. A Single Flux Quantum Cryogenic Random Access Memory. *Extended Abstracts of 7th International Superconducting Electronics Conference (ISEC'99)* 124–127 (Proc. Ext. Abstracts 7th Int. Supercond. Electron. Conf. (ISEC'99), 1999).
45. Braiman, Y., Neschke, B., Nair, N., Imam, N. & Glowinski, R. Memory states in small arrays of Josephson junctions. *Phys. Rev. E* (2016) doi:10.1103/PhysRevE.94.052223.
46. Nair, N., Jafari-Salim, A., D'Addario, A., Imam, N. & Braiman, Y. Experimental demonstration of a Josephson cryogenic memory cell based on coupled Josephson junction arrays. *Supercond. Sci. Technol.* **32**, 115012 (2019).
47. Collaudin, B. & Rando, N. Cryogenics in space: A review of the missions and of the technologies. *Cryogenics (Guildf).* (2000) doi:10.1016/S0011-2275(01)00035-2.

# Supplementary Materials: A Non-Volatile Cryogenic Random-Access Memory Based on Quantum Anomalous Hall Effect

## 1. Phenomenological Model of Quantum Anomalous Hall States in tBLG Moiré Heterostructure

We have developed a Verilog-A based phenomenological model for the recently observed intrinsic quantum anomalous Hall effect (QAHE) in tBLG moiré heterostructure, and demonstrate the operation of the proposed QAHE-based memory device [1]. In the main text, we have shown the DC bias current dependence of Hall resistance obtained from our model. Here, in the supplementary Fig. S1, we highlight a few additional device characteristics obtained from the model. In Figs. S1(a) & (b), we show the AC bias current dependence of Hall resistance [Fig. S1(b)] at 4 K temperature where a 40 nA$_{p-p}$ AC bias current [Fig. S1(a)] has been applied. Another feature of our model is the incorporation of the temperature dependence of the Hall resistance. To demonstrate the temperature dependence of the Hall resistance, in Fig. S1(c) we show the DC bias current dependence of the Hall resistance at different temperatures obtained from our model.

## 2. Feasibility of the Approach Used for Write and Read Operation

In the main text, we have presented the read and write operation for a cell in a 3D cross-point memory array. During the read operation from a cell, the Hall voltage terminals of the neighboring cells in the same row are electrically connected in series. Then, we use the algebraic sum of the Hall voltages of the cells in a row ($V_{hr-i} = \sum_{j=1}^{n} V_{xy-ij}$) to sense the memory state stored in a particular cell of that row. Note that positive/negative Hall voltage corresponds to logic '0'/'1' for the read operation. Notably, during the memory operations, it is important that *(i)* the sign of $V_{hr}$ remains same as that of the Hall voltage of the accessed cell, and *(ii)* the current that flows through the half-accessed and unaccessed cells during read and write operations do not affect the memory states stored in those cells. Figures S2(d) and S2(e) depict the repeated write operations (write 0 → 0 and 1 → 1) of the accessed cell. It is also clear from these figures that the Hall resistance of the other cells remain undisturbed. We also show the Hall voltages across the half-accessed and unaccessed cells during write '0' [Fig. S2(f)], write '1' [Fig. S2(g)], and read operation when logic '1' [Fig. S2(h)] and logic '0' [Fig. S2(i)] are stored in the half-accessed and unaccessed cells. As seen in these figures, the Hall voltages across the half-accessed row (HAR) cells are in nano-volt range during all the operations. Moreover, in Fig. 4(k) of the main text, we show the cell bias currents through all the cells during different memory operations. The choice of the bias current value further ensures that the stored memory sates in the HAR, half-accessed column (HAC) and unaccessed cells are not disturbed during read and write operations in a particular accessed cell.

## 3. Comparison with the State-of-the-art Cryogenic Memories:

Finding a low-power-consuming, high-capacity cryogenic memory is one of the major challenges in implementing a scalable quantum computer [2]. A number of designs based on Josephson junctions and CMOS architecture have been proposed. However, they have critical bottlenecks. While the Josephson junction-based memories are fast and energy efficient, they suffer from low capacity [3,4]. Conversely, CMOS memories lack speed and energy efficiency, albeit they have high capacity. To address this issue, one may think of hybrid memories utilizing the advantages of both the technology [5–8]. Intriguingly, this approach still poses a challenge of finding suitable interface circuits. Our proposed QAHE based memory, on the other hand, shows a strong potential as a cryogenic memory platform in quantum computers and cryogenic electronic systems thanks to its very small area and very low power operations. Moreover, cross-point array structure that we have constructed for this memory bestows the advantage of the high-density random-access memory. In Table S1, we summarize the comparison between our proposed memory and the state-of-the-art cryogenic memories in terms of area and power consumption. Our proposed QAHE based non-volatile memory offers at least a million times reduction in the cell area and a 1000 times reduction in the cell read/write power compared to the state-of-the-art cryogenic memory devices.

**Table S1: Comparison with the state-of-the-art Cryogenic Memories**

| Cryogenic Memories | Cell Area | Power Consumption |
|---|---|---|
| Superconducting Memory Chip [9] | $50 \times 46$ µm$^2$ | 1.62 mW for 2Kb RAM [Approximated 0.79 µW per cell] |
| VT RAM [10] | $55 \times 55$ µm$^2$ | 6.7 mW for 4 Kb RAM [Approximated 1.64 µW per cell] |
| SFQ CRAM [11] | $40 \times 45$ µm$^2$ | 2.4 mW for 16 Kb RAM [Approximated 0.15 µW per cell] |
| High frequency RAM [4] | $55 \times 55$ µm$^2$ | Not Reported |
| Hybrid Josephson-CMOS Memory [6,7] | $6.5 \times 7.5$ µm$^2$ | 0.7 mW for reading and 1.4 mW for writing for 64 Kb RAM [Approximated 0.01 µW for reading and 0.02 µW for writing per cell] |
| Small Array of Josephson Junctions [12,13] | $5 \times 5$ mm$^2$ (Chip Size) | ~ 5 nW (maximum) for reading and writing per cell (predicted) |
| QAHE based Proposed Memory (This work) | 130 nm$^2$ | 0.96 pW for writing and 0.1 pW for reading per cell (simulated) |

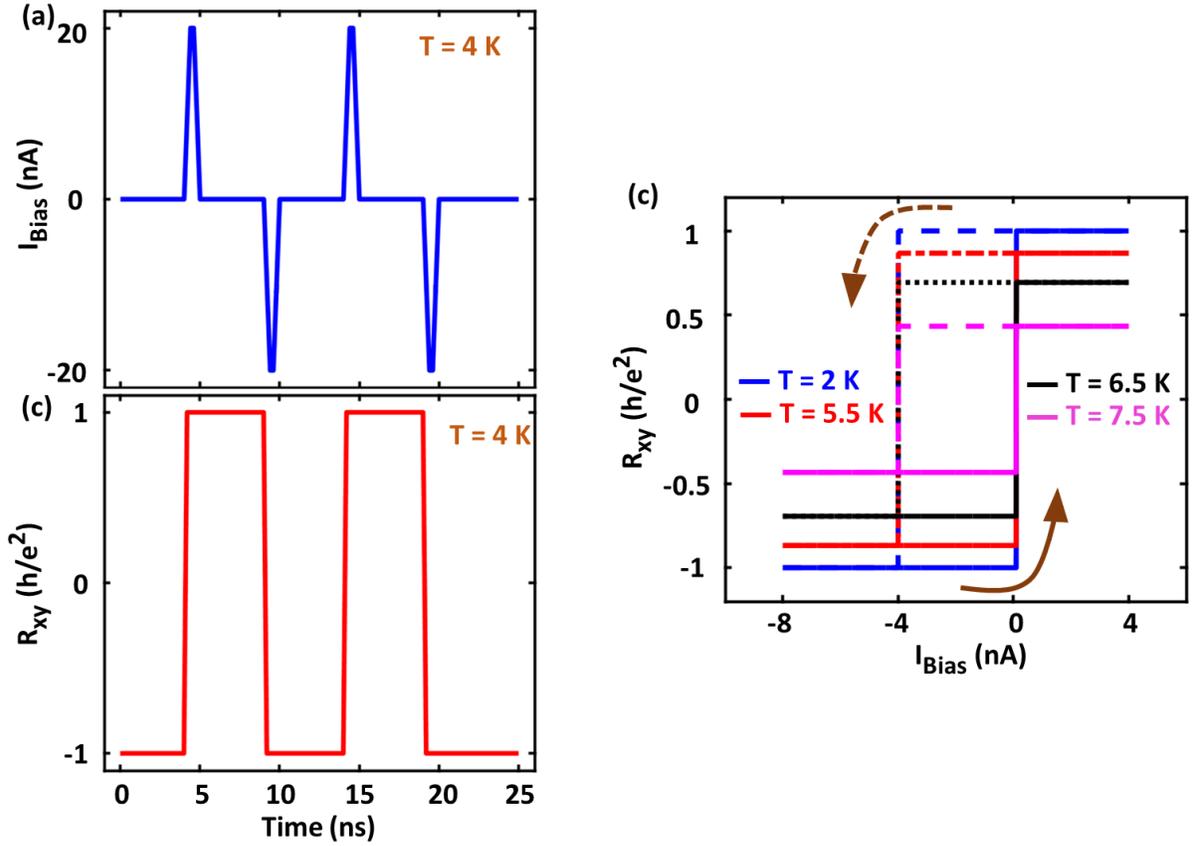

**Fig. S1: Time dynamics of the bias current and the Hall resistance and the temperature dependence obtained from our developed phenomenological model for the studied QAHE device.** Time dynamics of **(a)** the bias current, $I_{Bias}$ and **(b)** the resulting Hall resistance, $R_{xy}$ in tBLG moiré heterostructure at 4 K temperature. **(c)** Hall resistance, $R_{xy}$ as a function of the bias current, $I_{Bias}$ at different temperatures obtained from the phenomenological model.

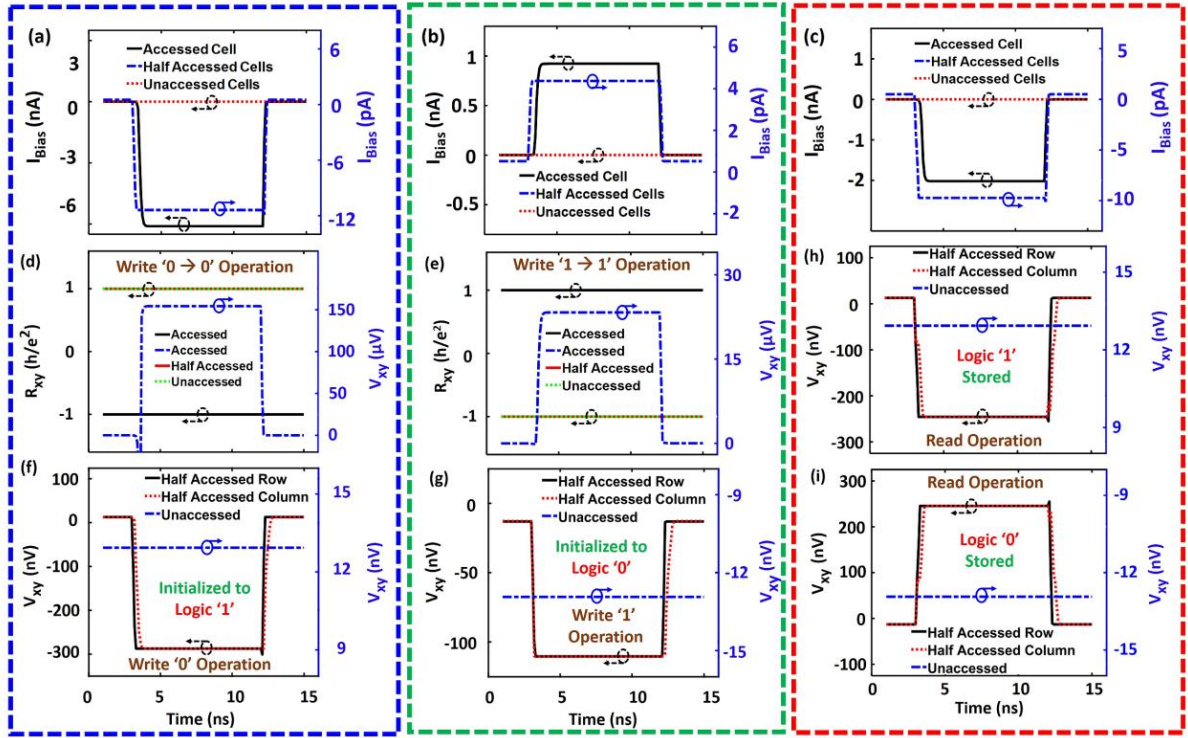

**Fig. S2: Additional details for the read/write operations.** Cell Bias currents ($I_{Bias}$) corresponding to the choice of specific values of $V_{ACC}$ during **(a)** write 0', **(b)** write '1', and **(c)** read operations. Time dynamics of the Hall resistance ($R_{xy}$) and Hall voltage ($V_{xy}$) across the accessed cell during the '0 → 0' write operation. To examine the possibility of accidental write operations occurring in other cells, we consider that all other cells are storing logic '1' ($R_{xy} = h/e^2$). Clearly, the $R_{xy}$ remains stable not only for the accessed cell, but also for all other cells, as expected. (e) Similar argument applies to the '1→1' write operation in the accessed cell. The $V_{xy}$ levels for the half-accessed and unaccessed cells are limited to sub-microvolts for (f) write '0' and (g) write '1' operations. Although $V_{xy}$ do not play any role in the memory write mechanism, these residual voltage levels lead to mild leakage power loss. Finally, we present the residual $V_{xy}$ levels that appear across the read terminals of half-accessed and unaccess cells during the (h) read '0' and (i) read '1' operations. Such ultra-low $V_{xy}$ levels ensure low leakage power throughout the array.

**References:**

1. Serlin, M. *et al.* Intrinsic quantized anomalous Hall effect in a moiré heterostructure. *Science (80-. ).* (2020) doi:10.1126/science.aay5533.
2. Tolpygo, S. K. Superconductor digital electronics: Scalability and energy efficiency issues. *Low Temperature Physics* (2016) doi:10.1063/1.4948618.
3. Polonsky, S. V., Kirichenko, A. F., Semenov, V. K. & Likharev, K. K. Rapid Single Flux Quantum Random Access Memory. *IEEE Trans. Appl. Supercond.* (1995) doi:10.1109/77.403223.
4. Nagasawa, S., Numata, H., Hashimoto, Y. & Tahara, S. High-frequency clock operation of josephson 256-word x 16-bit rams. *IEEE Trans. Appl. Supercond.* (1999) doi:10.1109/77.783834.
5. Ghoshal, U., Kroger, H. & Van Duzer, T. Superconductor-Semiconductor Memories. *IEEE Transactions on Applied Superconductivity* (1993) doi:10.1109/77.233542.
6. Liu, Q. *et al.* Latency and power measurements on a 64-kb hybrid Josephson-CMOS memory. in *IEEE Transactions on Applied Superconductivity* (2007). doi:10.1109/TASC.2007.898698.
7. Feng, Y. J. *et al.* Josephson-CMOS hybrid memory with ultra-high-speed interface circuit. in *IEEE Transactions on Applied Superconductivity* (2003). doi:10.1109/TASC.2003.813902.
8. Yau, J. B., Fung, Y. K. K. & Gibson, G. W. Hybrid cryogenic memory cells for superconducting computing applications. in *2017 IEEE International Conference on Rebooting Computing, ICRC 2017 - Proceedings* (2017). doi:10.1109/ICRC.2017.8123684.
9. Yuh, P. F. A 2-kbit Superconducting Memory Chip. *IEEE Trans. Appl. Supercond.* **3**, 3013–3021 (1993).
10. Tahara, S. *et al.* 4-Kbit Josephson Nondestructive ReadOut Ram Operated At 580 psec and 6.7 mW. *IEEE Trans. Magn.* (1991) doi:10.1109/20.133751.
11. Kirichenko, A., Mukhanov, O., Kirichenko, A. F., Mukhanov, O. A. & Brock, D. K. A Single Flux Quantum Cryogenic Random Access Memory. *Extended Abstracts of 7th International Superconducting Electronics Conference (ISEC'99)* 124–127 (Proc. Ext. Abstracts 7th Int. Supercond. Electron. Conf. (ISEC'99), 1999).
12. Braiman, Y., Neschke, B., Nair, N., Imam, N. & Glowinski, R. Memory states in small arrays of Josephson junctions. *Phys. Rev. E* (2016) doi:10.1103/PhysRevE.94.052223.
13. Nair, N., Jafari-Salim, A., D'Addario, A., Imam, N. & Braiman, Y. Experimental demonstration of a Josephson cryogenic memory cell based on coupled Josephson junction arrays. *Supercond. Sci. Technol.* **32**, 115012 (2019).